\documentclass[conference]{IEEEtran}
\IEEEoverridecommandlockouts
\usepackage{cite}
\usepackage{amsmath,amssymb,amsfonts}
\usepackage{algorithmic}
\usepackage{graphicx}
\usepackage{textcomp}
\usepackage{xcolor}
\usepackage{subcaption}
\usepackage{amsmath}
\def\BibTeX{{\rm B\kern-.05em{\sc i\kern-.025em b}\kern-.08em
    T\kern-.1667em\lower.7ex\hbox{E}\kern-.125emX}}
\begin{document}

\title{Pinpointing the Memory Behaviors of DNN Training\\
\thanks{This work is supported by the National Key R\&D Program of China~(under Grant No. 2017YFB1003103) and the Science Fund for Creative Research Groups of the National Natural Science Foundation of China~(under Grant No. 61521092).}
}

\author{\IEEEauthorblockN{Jiansong Li$^{\dagger\ddagger}$, Xiao Dong$^{*}$, Guangli Li$^{\dagger\ddagger}$, Peng Zhao$^\P$, Xueying Wang$^{\dagger\ddagger}$, Xiaobing Chen$^{\dagger\ddagger}$,\\Xianzhi Yu$^\P$, Yongxin Yang$^{\dagger\ddagger}$, Zihan Jiang$^{\dagger\ddagger}$, Wei Cao$^{\dagger}$, Lei Liu$^{\dagger}$, Xiaobing Feng$^{\dagger\ddagger}$}
  \IEEEauthorblockA{\textit{$^{\dagger}$State Key Laboratory of Computer Architecture, Institute of Computing Technology, CAS, Beijing, China}\\
  \textit{$^{\ddagger}$University of Chinese Academy of Sciences, Beijing, China}\\
  \textit{$^{*}$Youtu Lab, Tencent, Shanghai, China}
  \textit{$^{\P}$Huawei Technology Co., Ltd, Beijing, China}\\
  \{lijiansong, liguangli, wangxueying, chenxiaobing, yangyongxin, jiangzihan, caowei, liulei, fxb\}@ict.ac.cn\\
  devandong@tencent.com, zhaopeng71@huawei.com, yuxianzhi@huawei.com
  }
}

\maketitle

\begin{abstract}
The training of deep neural networks~(DNNs) is usually memory-hungry due to the limited device memory capacity of DNN accelerators. Characterizing the memory behaviors of DNN training is critical to optimize the device memory pressures. In this work, we pinpoint the memory behaviors of each device memory block of GPU during training by instrumenting the memory allocators of the runtime system. Our results show that the memory access patterns of device memory blocks are stable and follow an iterative fashion. These observations are useful for the future optimization of memory-efficient training from the perspective of raw memory access patterns.
\end{abstract}


\section{Introduction}

Nowadays, DNNs trend to go deeper and wider for higher accuracy~\cite{openai-20-gpt-3,google-21-switch-transformers}. The statistics from OpenAI show that the computation and memory requirements of DNNs are doubled every 3.4 months in recent years~\cite{openai-trend}. However, the memory capacity constraint of DNN accelerators limits the upper bound on the size of DNNs that can be offloaded to and trained. For instance, the DRAM size of Nvidia GPU with the latest Ampere architecture is 40GB~\cite{nvidia-a100-gpu}. Even such a large memory capacity fails to satisfy the demand of some DNNs. For example, the typical Inception-V4~\cite{aaai-17-inceptionv4} requests up to 45GB of device memory to keep the entire DNN on the GPU during training~\cite{hpdc-20-nonlinear}. It is important to reduce the memory pressures of DNN training. Characterizing the memory behaviors of DNN training is critical to optimize the memory usage of DNN workloads.

\section{Benchmarking Method}

To work around the memory pressures of DNN training, we carefully pinpoint the memory behaviors~(including malloc, free, read, write) of each device memory block by manually instrumenting the memory allocators of PyTorch's runtime system. We firstly conduct a case study of the trivial MLP to show the memory access patterns of DNN training. To see where the device memory is spent and which kinds of memory content is the bottleneck, we also make a detailed memory occupation breakdown for the typical linear and non-linear~\cite{hpdc-20-nonlinear} DNNs with different batch size and layer structures.

\begin{figure}[!htbp]
\vspace{-0.45cm}
 \centering
 \includegraphics[width=0.40\textwidth]{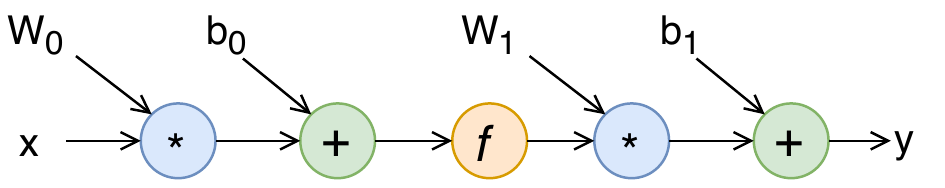}
 \vspace{-0.15cm}
 \caption{\label{fig:mlp-case-study} Layer topology of MLP. Note that star means the mat\_mul operator, plus is add\_bias, $f$ is ReLU activation.}
\end{figure}

\begin{figure}[!htbp]
 \centering
 \vspace{-0.85cm}
 \includegraphics[width=0.50\textwidth]{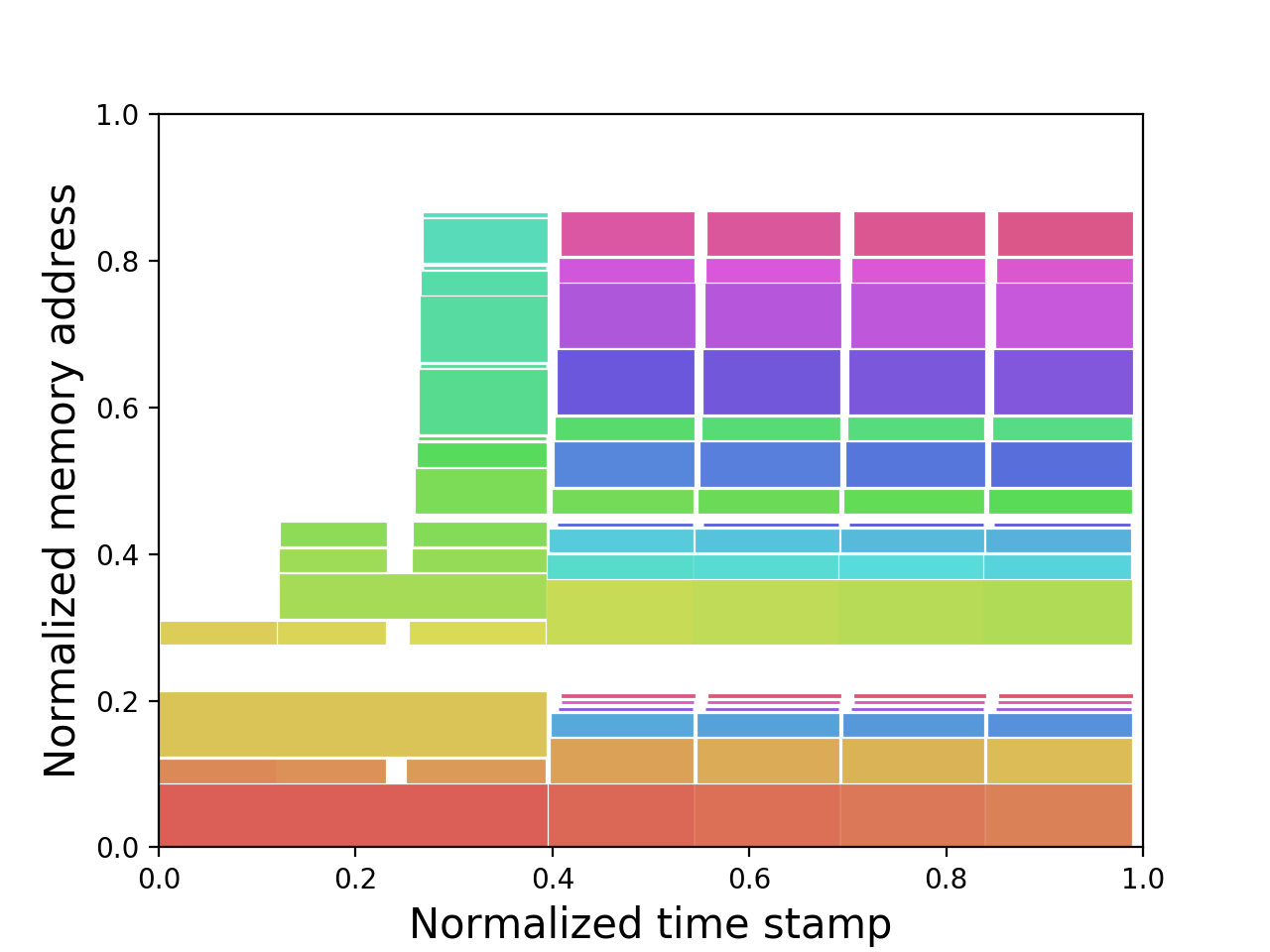}
 \vspace{-0.25cm}
 \caption{\label{fig:mlp-gantt} Gantt chart of the first five iterations in MLP training on the Nvidia Titan X Pascal GPU.}
\end{figure}

\section{Experimental Results}

\noindent
\textbf{Iterative Memory Access Patterns}. Due to space limit, we only use a trivial MLP to show the memory access patterns of DNN training~(our observations are also appliable to other DNNs). Its layer topology is shown in Fig.~\ref{fig:mlp-case-study}. The shape of $W_0$ is~(2, 12288), $b_0$:~(12288), $W_1$:~(12288, 2), $b_1$:~(2). In Fig.~\ref{fig:mlp-gantt}, each rectangle represents the memory access to a certain device memory block of GPU. Width of the rectangle is the elapsed time from the allocation time to the free time of current device memory block, which can be regarded as the lifetime of current memory block. Height of the rectangle denotes the size of current device memory block. We observe that there are obvious iterative memory access patterns in the first five rounds of MLP training. Besides, the overlap between rectangles along the y-axis represents the overlap of the live ranges of different memory blocks. The blank space between two rectangles along the y-axis represents the memory fragments of GPU. We can also observe that there are fewer memory fragments during MLP training.

\begin{figure}[!htbp]
 \centering
 \begin{subfigure}[c]{.15\textwidth}
    \centering
    \includegraphics[width=1\linewidth]{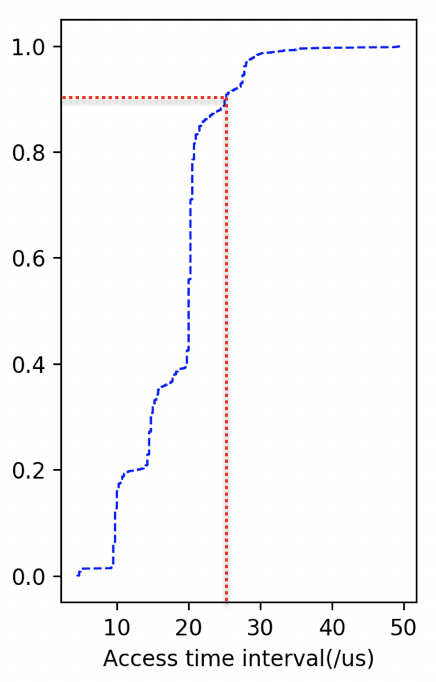}
    \caption{CDF~\cite{1989-cdf}.}
    \label{fig:mlp-cdf}
 \end{subfigure}
 \begin{subfigure}[c]{.31\textwidth}
    \centering
    \includegraphics[width=1\linewidth]{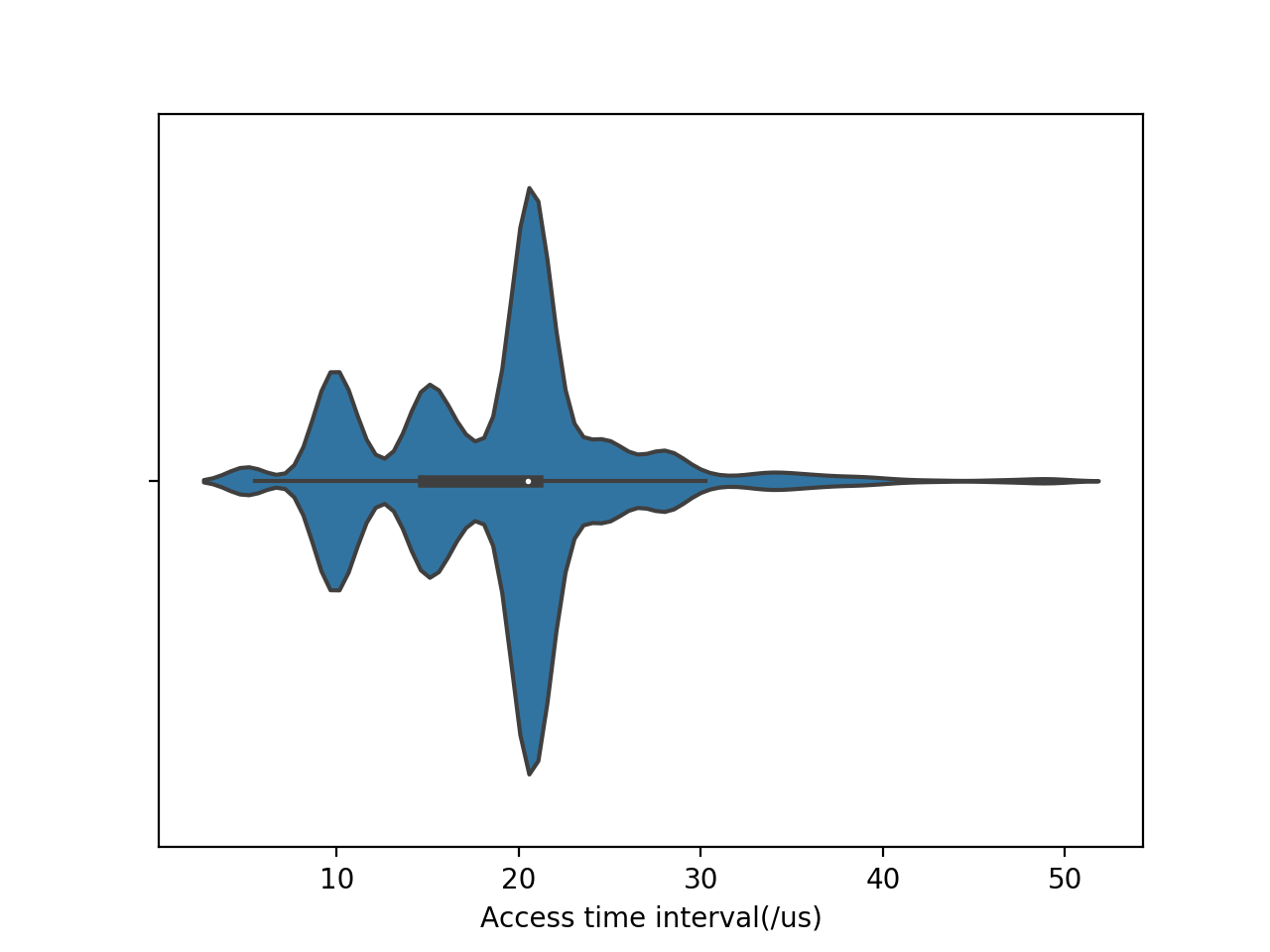}
     \caption{Violin plot~\cite{violin-plot}.}
    \label{fig:mlp-violinplot}
  \end{subfigure}
  \vspace{-0.25cm}
  \caption{\label{fig:mlp-cdf-violinplot} CDF and violin plot of the memory block access intervals in MLP training on the Nvidia Titan X Pascal GPU.}
\end{figure}

Furthermore, Fig.~\ref{fig:mlp-cdf-violinplot} shows the distributions of these memory behaviors. The access time interval~(abbr $ATI$) is the elapsed time between two adjacent memory access to the same memory block. As presented in Fig.~\ref{fig:mlp-violinplot}, the $ATI$s of most memory behaviors range from 10us to 25us, and their distributions are relatively concentrated. Besides, Fig.~\ref{fig:mlp-cdf} shows that the $ATI$s of 90\% of the memory behaviors are less than 25us. Suppose we want to reduce the device memory pressure by swapping data back and forth between the host CPUs and the device GPUs~\cite{asplos-20-swapadvisor,21-zerooffload}, the $ATI$ of current device memory block is $T$. Assume the memory bandwidth from device GPU to host CPU is $B_{d2h}$ and that of host to device is $B_{h2d}$. Then the maximum value of memory swapping size $S$ without sacrificing the runtime performance should meet the following equation:
\begin{equation}
\label{max-swapping-size}
\begin{aligned}
    \frac{S}{B_{d2h}} + \frac{S}{B_{h2d}} \leq T, \\
    S \leq \frac{T}{1/B_{d2h} + 1/B_{h2d}}
 \end{aligned}
\end{equation}

We measured the \textit{memcpy} bandwidth between the device GPU and host CPU with the \textit{bandwidthTest} tool from CUDA SDK samples~\cite{cuda-sdk-samples}. The pinned memory transfer bandwidth from host to the device is 6.3GB/s, from device to host is 6.4GB/s. So the maximum value of memory swapping size without sacrificing the runtime performance is about 79.37KB~(by equation~\ref{max-swapping-size}, $S \leq \frac{25us}{1/6.4GB/s+1/6.3GB/s}=79.37KB$). This is just a drop in the bucket for reducing the memory footprint of MLP training. It seems that memory swapping is promising less.
\begin{figure*}[!htbp]
 \centering
 \includegraphics[width=0.95\textwidth]{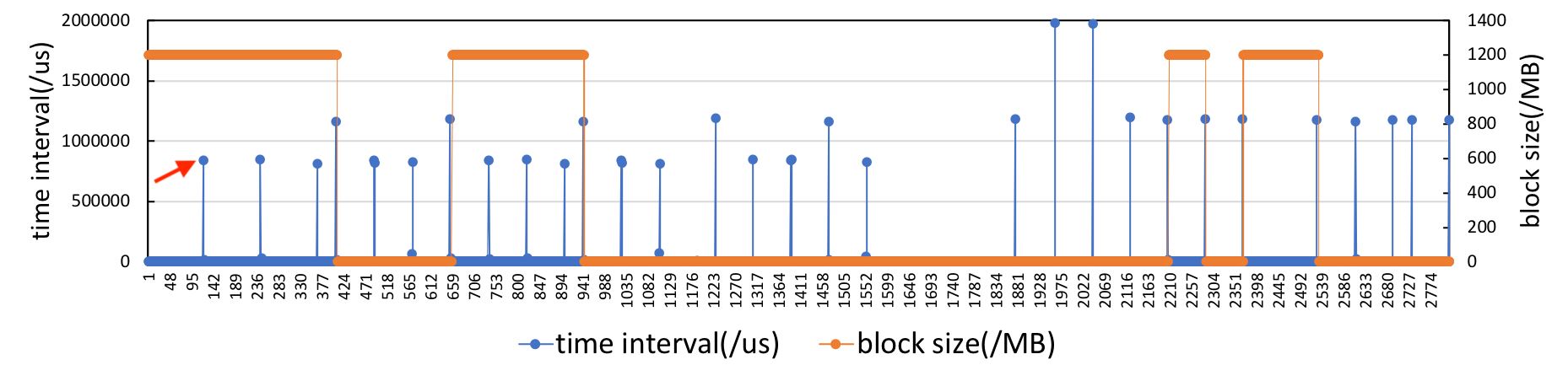}
 \vspace{-0.25cm}
 \caption{\label{fig:mlp-detail} Pair-wise $ATI$ and the corresponding device memory block size of each memory behavior during MLP training on the Nvidia Titan X Pascal GPU. Note that the x-axis is the index of each memory behavior.}
\end{figure*}

The detailed memory behaviors of Fig.~\ref{fig:mlp-detail}, including $ATI$s along with the corresponding memory block size in MLP training, show that the $ATI$s of most memory access behaviors are negligible, which is consistent with the observation from Fig.~\ref{fig:mlp-cdf-violinplot}. However, there are some outliers whose access time interval is larger than 0.8s and the corresponding memory block size is larger than 600MB. For example, the $ATI$ and the corresponding memory block size of the red marked outlier is 840211us and 1200MB, respectively. In this case, the boundary data swapping size without sacrificing the runtime performance between the host and device is about 2.54GB~(by equation~\ref{max-swapping-size}, $S \leq \frac{0.8s}{1/6.4GB/s+1/6.3GB/s}=2.54GB$), which is much larger than 1200MB. This indicates that these outlier memory behaviors with high $ATI$s and large memory block size ought to be the focus of attention. They are the major contributors in terms of reducing the memory pressure of DNN training. In the future, we plan to build an automatic cost model to sift out these memory access behaviors to reduce the device memory pressure during training.

\begin{figure}[!htbp]
 \centering
 \vspace{-0.75cm}
 \includegraphics[width=0.35\textwidth]{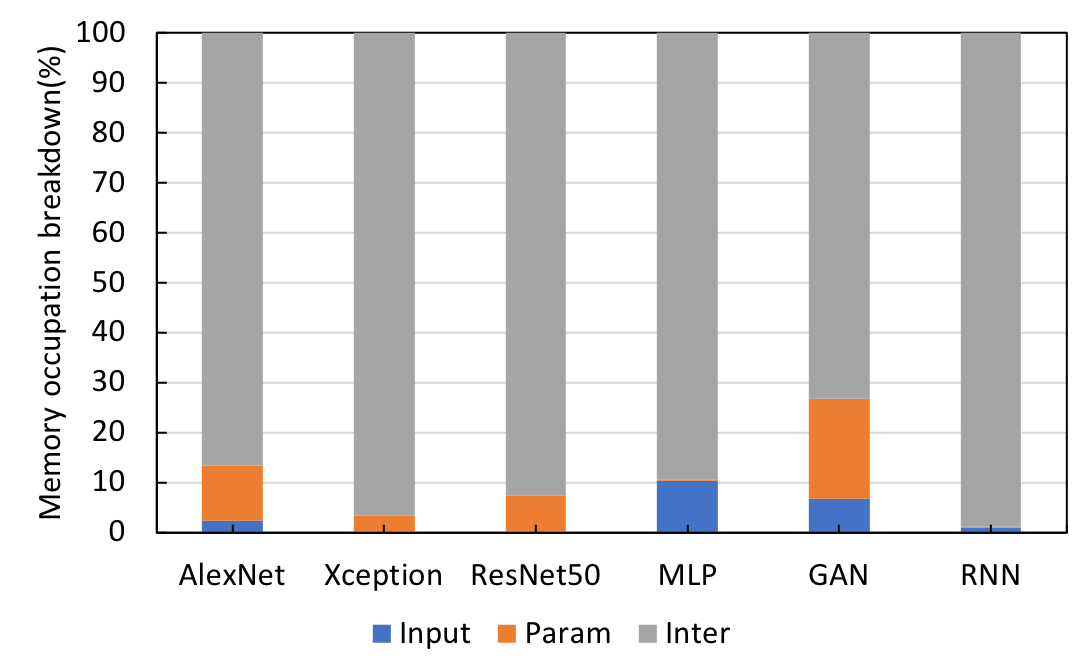}
 \vspace{-0.25cm}
 \caption{\label{fig:mem-breakdown} Memory occupation breakdown of typical DNN training on the Nvidia Titan X Pascal GPU.}
\end{figure}

\begin{figure}[!htbp]
\centering
\vspace{-0.5cm}
  \begin{subfigure}[c]{.23\textwidth}
    \centering
    \includegraphics[width=1\linewidth]{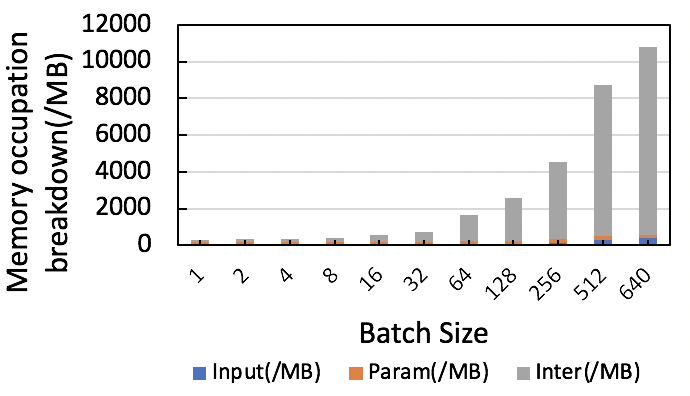}
    \caption{CIFAR-100~(32x32).}
    \label{fig:alexnet-mem}
  \end{subfigure}
  \begin{subfigure}[c]{.23\textwidth}
    \centering
    \includegraphics[width=1\linewidth]{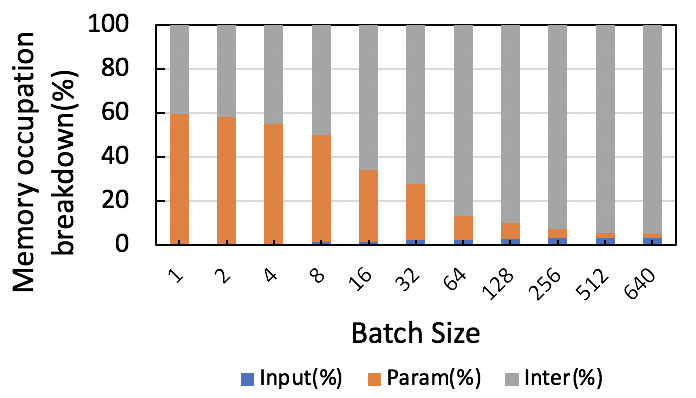}
    \caption{CIFAR-100~(32x32).}
    \label{fig:alexnet-mem-ratio}
  \end{subfigure}
  \vspace{-0.25cm}
 \caption{\label{fig:alexnet-mem-breakdown} Memory ocupation breakdown of linear DNN~(AlexNet) with different batch size.}
\end{figure}

\begin{figure}[!htbp]
\vspace{-0.5cm}
\centering
  \begin{subfigure}[c]{.245\textwidth}
    \centering
    \includegraphics[width=1\linewidth]{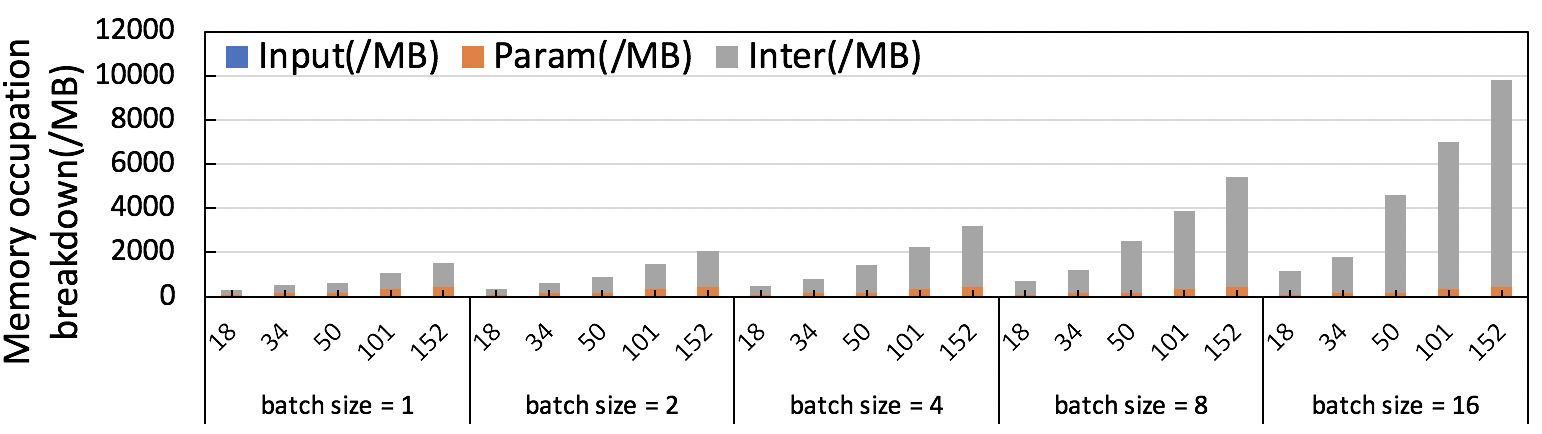}
    \caption{ImageNet~(224x224).}
    \label{fig:resnet-mem}
  \end{subfigure}
  \begin{subfigure}[c]{.23\textwidth}
    \centering
    \includegraphics[width=1\linewidth]{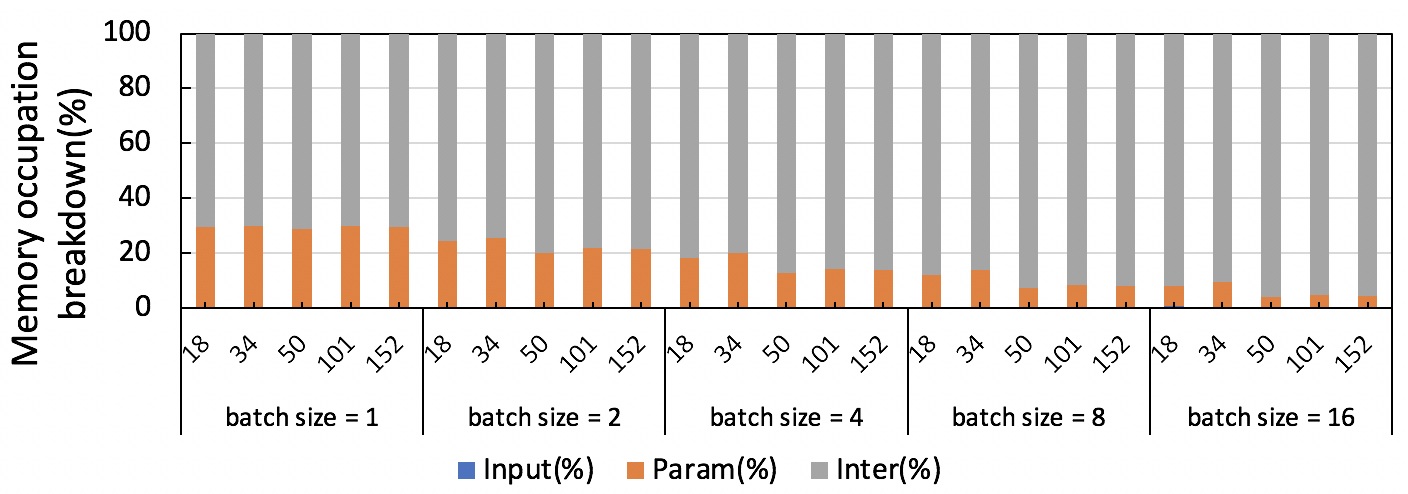}
    \caption{ImageNet~(224x224).}
    \label{fig:resnet-mem-ratio}
  \end{subfigure}
  \vspace{-0.25cm}
 \caption{\label{fig:resnet-mem-breakdown} Memory ocupation breakdown of non-linear DNN~(ResNet) with different layer structures~(ResNet18, 34, 50, 101, 152).}
\end{figure}

\noindent
\textbf{Device Memory Occupation Breakdown}. During training, the memory consumption falls into three categories from the perspective of device memory storage contents, i.e., input data, parameters and intermediate results~\cite{1998-deeplearning}. We observe from Fig.~\ref{fig:mem-breakdown} that, for most DNNs, parameters only account for a small fraction of the total memory footprint during training. This indicates that weight pruning or quantization techniques are not efficient for reducing the memory pressures of DNN training~\cite{weight-redundant-nips-13,patdnn-asplos-20}. In training, the intermediate results are the primary contributor to the device memory footprint. To show the detailed effects of datasets, batch size and model layer structures over the memory footprint, we explored the enumeration space of these involving factors. Due to space limit, we only report the data of one linear DNN~(AlexNet~\cite{nips-12-imagenet}) and one non-linear DNN~(ResNet with different numbers of residual layer blocks~\cite{cvpr-16-resnet}) with CIFAR-100 and ImageNet dataset under different batch size. As shown in Fig~\ref{fig:alexnet-mem-breakdown}, for AlexNet, with the growth of batch size, those intermediate results gradually dominate the device memory consumption and the occupation of parameters is gradually weakened. Meanwhile, the impact of input data increases slightly. This observation is also applicable to the non-linear ResNet with different numbers of residual layer blocks~(Fig.~\ref{fig:resnet-mem-breakdown}).

\section{Conclusion \& Future Directions}

From our characterizing results, we highlight the following observations. The memory behaviors of DNN training illustrate obvious iterative patterns~(Fig.~\ref{fig:mlp-gantt}). Besides, the distributions of the $ATI$s of most memory behaviors are relatively concentrated. Especially, the $ATI$s of most memory access behaviors are negligible, which can be filtered out by the bandwidth between host and device memory~(Fig.~\ref{fig:mlp-cdf-violinplot}). Most importantly, those outlier memory behaviors with high $ATI$ and large memory block size ought to be the focus of attention~(Fig.~\ref{fig:mlp-detail}). They are the major contributors to reduce the memory pressure of DNN training.

Based on these observations, and inspired by the swapping-based memory pressure reduction schemes~\cite{asplos-20-swapadvisor,21-zerooffload}, we plan to propose a more general approach that takes the memory access patterns as input to automatically address the device memory pressure issues of DNN training with small runtime overhead.

\bibliographystyle{IEEEtran}
\bibliography{references}

\end{document}